# Manipulation of Two-Color Stationary Light Using Coherence Moving Gratings

S. A. Moiseev[1,2] and B. S. Ham[1]

[1]*The Graduate School of Information and Communications, Inha University, Incheon 402-751 S. Korea*
[2]*Kazan Physical-Technical Institute of Russian Academy of Sciences, Kazan 420029 Russia*
E-mails: moiseev@inha.ac.kr; bham@inha.ac.kr

We propose the dynamic control of two-color stationary light in a double-lambda four-level system using electromagnetically induced transparency. We demonstrate the complete localization of two-color quantum fields inside a medium using coherence moving gratings resulting from slow-light based atom-field interactions in backward nondegenerate four-wave mixing processes. The quantum coherent control of the two-color stationary light opens a door to deterministic quantum information science which needs a quantum nondemolition measurement, where the two-color stationary light scheme would greatly enhance nonlinearity with increased interaction time.

      For the last decade light-matter interactions using a coherent control has drawn much attention to the area of nonlinear quantum optics for the possibility of superior enhancement of nonlinearity which otherwise is intrinsically small and requires intense light [1]. Such interactions are most important for quantum information technologies which especially need repeated non-destructive measurements of physical observables [2]. Even though there have been many intensive studies of quantum nondemolition measurements, an intrinsically low nonlinear coefficient has been a major problem where a few photons are utilized [3]. To overcome such a drawback, a cavity-QED technique has been applied to achieve cavity-enhanced light intensity and cavity-lengthened interaction time [4]. However, the requirement of the atom-field interaction in a high-Q cavity gives rise to a problem in the use of this technique especially in the case of a traveling light scheme.

      Recently electromagnetically induced transparency (EIT) [5] has been successfully applied to nonlinear quantum information to obtain giant Kerr nonlinearity [6], quantum switching [7], quantum memory [8,9], quantum entanglement [10], and quantum computing [11]. EIT is a quantum optical phenomenon, where atom-field interactions modify the refractive index of the medium to be transparent even to a resonant light [12-14]. This type of refractive index change of the medium results in the ultra-slow group velocity of a light pulse observed recently [15,16]. In addition, highly efficient nondegenerate four-wave mixing generations have been demonstrated using EIT-induced giant Kerr nonlinearity [17], which is likely to provide broad potential for using a very weak optical field to obtain a $\pi$-phase shift [10,18,19], where a quantum superposition such as Schroedinger's cat can be realized [10,19].

      Very recently, stationary light using EIT-induced slow light in a $\Lambda$-type atomic medium has been observed with a standing-wave grating formation caused by the counter-propagating laser fields [20]. The interaction time of a traveling light with a localized atom is independent of the group velocity of the light, whereas a standing light can enormously increase the interaction time. This is the essence of a standing light in nonlinear quantum optics with a spatially limited atomic ensemble, and enhanced nonlinear effects would be expected owing to the lengthened photon-atom interaction time. Unlike a traveling slow-light scheme based on giant Kerr nonlinearity [6,10,17-19], the standing-light scheme can be similar to a cavity-QED. Thus, one can apply the standing light to deterministic quantum information science where quantum nondemolition measurement is essential and low nonlinearity is a crucial problem. Here, it should be noted that the standing light is completely different from the quantum memory phenomenon [8,9,16,21], where a complete quantum conversion process is required between photons and atoms and nonlinearity is not involved.

      In this paper, we present a quantum manipulation of a weak quantum field for two-color stationary light using coherence moving gratings based on backward nondegenerate four-wave mixing processes in a resonant double-$\Lambda$ scheme (see Fig. 1). Nonlinear enhancement of the double-$\Lambda$ scheme has been already demonstrated in both atomic media for a continuous wave scheme [22,23] and a solid medium for a pulsed scheme [24]. Recently quantum memory [9] and quantum entanglement [10] have been demonstrated using two-color slow lights in a double-$\Lambda$ scheme. The astonishing consequence of the

present two-color stationary light is in the potential applications to deterministic quantum information based on quantum nodemolition measurement due to enormously enhanced nonlinearity in the interactions with a localized atomic system.

Figure 1 shows an energy level diagram for the two-color stationary light based on a double-Λ scheme. As shown in Fig. 1, the control fields $\Omega_+$ and $\Omega_-$ have transition frequencies resonant from the state |2> to the excited states |3> and |4> with opposite propagation directions $\kappa_+$ and $\kappa_-$, respectively. As is well known in the double-Λ system, the propagation direction $\mathbf{k}_-$ of the field $E_-$ is determined by the phase matching with Bragg condition: $\mathbf{k}_-=\kappa_+-\mathbf{k}_++\kappa_-$. Unlike previous stationary light experiment based on standing-wave grating [20], the present two-color stationary light can utilize a single frequency traveling light to create two spectral components of stationary lights via a double-EIT scheme, in which both spectral components are absorption free [25].

For a theoretical analysis of the two-color stationary lights based on a double-Λ type four-level system in Fig. 1, we introduce a quantum fields $E_\sigma = \sqrt{\hbar\omega_\sigma/(2\varepsilon_o V)} A_\sigma e^{-i\omega_\sigma(t-\sigma z/c)} + H.C.$, which is traveling in a ±z direction, where $A_\sigma$ are slowly varying field operators, we assume that the quantization volume V=1, The field (for σ = +) is resonant with atomic transitions of |1> − |3> and |1> − |4> (for σ = −), where $\sigma = +/-$ stands for forward/backward field, respectively. In the interaction picture the Hamiltonian of the quantum fields with atoms driven by the control laser fields $\Omega_+(t)$ and $\Omega_-(t)$ is

$$H = \hbar g_+ \sum_{j=1} A_+(t,z_j) P_{31}^j e^{i\omega_+ z_j/c} + \hbar g_- \sum_{j=1} A_-(t,z_j) P_{41}^j e^{-i\omega_- z_j/c} - \hbar \sum_{j=1} \{\Omega_+ P_{32}^j \exp[i(Kz_j + \varphi_+)] + \Omega_- P_{42}^j \exp[-i(Qz_j - \varphi_-)]\} + H.C., \quad (1)$$

where $P_{nm}^j = (P_{mn}^j)^+$ are the atomic operators, $g_\sigma = \wp_\sigma \sqrt{\omega_\sigma/(2\varepsilon_o \hbar V)}$ is a coupling constant of photons with atoms, $\wp_\sigma$ is a dipole moment for each transition [1], $\varphi_{\sigma=\pm}$ are the phases of the control fields, $\omega_+ = \omega_{31}$, and $\omega_- = \omega_{41}$. Using Eq. (1) we derive equations for the field $\hat{A}_\sigma$ and atomic operators $P_{mn}^j$. Adding the decay constants $\gamma_4 = \gamma_3 = \gamma$ to coherence equations $P_{13}^j$ and $P_{14}^j$, and $\gamma_2$ to the spin coherence $P_{12}^j$, we introduce new slow-light field operators $\Psi_\sigma = e^{-i(\varphi_\sigma + \sigma\omega_{21}z/c)} \sqrt{N} g_\sigma A_\sigma / \Omega_\sigma$ (N is atomic density) subject to the adiabatic approach [9]. Assuming slowly varying amplitudes of the laser fields and ignoring the atomic population on the excited levels |3> and |4> due to weak field $E_{\sigma=\pm}$ under the slow-light propagation $v_g \cong c\Omega_+^2/Ng_+^2 \ll c$, we obtain the following coupled wave equations for the new operators:

$$(\tfrac{\partial}{\partial z} + i\tfrac{\omega_{21}}{c})\Psi_+(\tau,z) = -\xi_+ \alpha_-[\Psi_+(\tau,z) - \Psi_-(\tau,z)] - \tfrac{\partial}{c\partial\tau}\{\alpha_+\Psi_+(\tau,z) + \alpha_-\Psi_-(\tau,z)\} - (\gamma_2'/c)\Psi_+(\tau,z), \quad (2)$$

$$(\tfrac{\partial}{\partial z} - i\tfrac{\omega_{21}}{c})\Psi_-(\tau,z) = -\xi_- \alpha_+[\Psi_+(\tau,z) - \Psi_-(\tau,z)]$$
$$+ (g_-/g_+)^2 \tfrac{\partial}{c\partial\tau}\{\alpha_+\Psi_+(\tau,z) + \alpha_-\Psi_-(\tau,z)\} + (g_-/g_+)^2(\gamma_2'/c)\Psi_-(\tau,z), \quad (3)$$

where $\xi_\sigma = Ng_\sigma^2/c\gamma$ are the absorption coefficients, $\alpha_\sigma = \Omega_\sigma^2/[\gamma\gamma_2 + \Omega_\Sigma^2]$, $\Omega_\Sigma^2 = \Omega_+^2 + \Omega_-^2$, $\tau = \int_{-\infty}^t dt' [\gamma\gamma_2 + \Omega_\Sigma^2]/(g_+^2 N)$ is a new time scale and $\gamma_2' = Ng_+^2[\gamma\gamma_2 + \Omega_\Sigma^2]^{-1}\gamma_2$.

If the counter-propagating control field is turned off, $\Omega_-=0$ the Eqs. (2) and (3) satisfy slow-light wave equations. We note that Eqs. (2) and (3) coincide with the standing single-frequency light based on a standing wave grating in a three-level system if $g_+ = g_-$ with Doppler broadening [20]. It should be noted that Eqs. (2) and (3) shows universal coupled equations of stationary lights whether the active medium is Doppler broadened or not (will be discussed elsewhere [26]). In Eqs. (2) and (3) we ignore the terms with $\omega_{21}/c$ assuming the frequency splitting $\omega_{21}$ is sufficiently small to satisfy phase matching for the interaction between the fields $\Psi_+$ and $\Psi_-$ [27].

Using the Fourier transformation $\Psi_\sigma(\tau,z) = \int_{-\infty}^\infty dk e^{ikz} \widetilde{\Psi}_\sigma(\tau,k)$, we find the analytical solution of Eqs. (2) and (3):
$$\Psi_\sigma(\tau,z) = \int_{-\infty}^\infty dk \exp\{+ikz + i\int_{\tau_o}^\tau d\tau' \omega(\tau',k)\} \widetilde{\Psi}_\sigma(\tau_o,k), \quad (4)$$

$$\omega(\tau',k) = i\eta\gamma_2' - \frac{ck\{\eta(\xi_-\alpha_+ - \xi_+\alpha_-) - ik\} - i(\partial/\partial\tau')(\eta^{-1}\xi_- - ik\tilde{\alpha})}{\{\eta^{-1}\xi_- - ik\tilde{\alpha}\}}, \quad (5)$$

where, $\tilde{\Psi}_-(\tau_o,k) = \tilde{f}_{-+}(k)\tilde{\Psi}_+(\tau_o,k)$, $\tilde{f}_{-+}(k) = (1+ik/\xi_+)/(1-ik/\xi_-)$, $\eta = (\Omega_\Sigma^2 + \gamma\gamma_2)\Omega_\Sigma^{-2}$ and $\tilde{\alpha} = \alpha_+ - (g_-/g_+)^2\alpha_-$. The functions $\tilde{\Psi}_{+,-}(\tau_o,z)$ are determined from the initial conditions for the field $A_{+,-}$ which enters the medium. Using Eq. (4) we get the coupled fields $\Psi_+(t,z')$ and $\Psi_-(t,z)$ expressed by nonlocal spatial relations:

$$\Psi_+(t,z) = \int_{-\infty}^{z+\varepsilon} dz' f_+(z-z')\Psi_-(t,z'), \quad (6\text{-}1)$$

$$\Psi_-(t,z) = \int_{z-\varepsilon}^{\infty} dz' f_-(z-z')\Psi_+(t,z'), \quad (6\text{-}2)$$

where $f_\sigma(z-z') = \xi_\sigma(1+\frac{\xi_\sigma}{\xi_{\sigma'}})\eta(z'-z)e^{-\xi_\sigma(z'-z)} - \frac{\xi_\sigma}{\xi_{\sigma'}}\delta(z-z')$ ($\sigma = \pm 1, \sigma' = \mp 1$, $\eta(x\geq 0)=1, \eta(x<0)=0$ and $\varepsilon$ is a small value $\varepsilon \to 0$). Thus, the field $\Psi_+(t,z)$ is correlated with $\Psi_-(t,z')$, where their spatial points (z<z') are within the correlation size $\xi_\sigma^{-1}$. This means that the field $\Psi_-(t,z)$ is a copy of the tuned field $\Psi_+(t,z')$ or vice-verse if the spatial size $l$ of the $\Psi_\pm(t,z')$ is larger than $\xi_\sigma^{-1}$. It is obvious that Eq. (6) should determine the spatial correlations of the quantum fields $E_+$ and $E_-$. Below we only focus our attention on the temporal dynamics of the fields where the field $A_+$ enters the medium when the control field $\Omega_+$ ($\Omega_-=0$) is turned on. Let the pulse $A_+$ have a Gaussian shape ($A_+(t,z=0) = A_{+,o}\exp\{-t^2/(2T^2)\}$), where $A_{+,o}$ and $T$ are the amplitude and temporal duration of the field. The solution of Eq. (4) describes the propagation of the field $\Psi_+$, that is, $A_+ = (g_+^2 N)^{-1/2}\Omega_+\Psi_+ e^{i(\varphi_+ + \omega_{21}z/c)}$ with a slow group velocity $v_g \cong \frac{c\Omega_+^2(0)}{Ng_+^2} \ll c$, and the initial spatial size $l_o = v_g T$ with an amplitude decay in accord with $\exp\{-\eta\gamma_2 t\}$. Thus, the field $A_+$ is transparent to the optically dense medium ($\xi_{+,-}l_o \gg 1$) owing to EIT.

Using the dispersion relation $\omega(k)$ in Eq. (5) we find the group velocity $v$ of the coupled light for the constant two control fields $\Omega_\sigma$ $v = -[(\Omega_\Sigma^2 + \gamma\gamma_2)/Ng_+^2]\partial\omega/\partial k|_{k=0} = c\eta^2[g_-^2\Omega_+^2 - g_+^2\Omega_-^2]/[N(g_+g_-)^2]$. Thus it is seen that the group velocity of the two color light can be completely stopped ($v=0$) if the stationary light condition $\Omega_+/g_+ = \Omega_-/g_-$ is satisfied. Ignoring small terms proportional to $ck^3/\xi_-^2$ in Eq. (5) we find the approximate solution of the integral (Eq. (4)) for arbitrarily varying amplitudes of the control laser fields $\Omega_\sigma$:

$$A_\sigma(\tau,z) = \frac{l_o}{cB(\tau)}\frac{\eta(\tau)(\Omega_\sigma(\tau)g_+)}{\eta(0)(\Omega_+(0)g_\sigma)}A_{+,o}\exp\{-\int_0^\tau(\eta\gamma_2')d\tau' - (\beta_\sigma(\tau) - z/c)^2/(2B^2(\tau))\}\exp\{i(\varphi_\sigma + \sigma\omega_{21}z/c)\}, \quad (7)$$

where

$$\beta_+(\tau) = \int_0^\tau d\tau' M_1(\tau'), \quad \beta_- = \beta_+ - z_o/c, \quad M_1 = \xi_-^{-1}\{\eta^2(\xi_-\alpha_+ - \xi_+\alpha_-) - \frac{\partial}{cd\tau'}(\eta\tilde{\alpha})\}, \quad (8)$$

$$B(\tau) = \sqrt{(l_o/c)^2 + (2/c)\int_0^\tau d\tau' M_2(\tau')}, \quad M_2 = \xi_-^{-2}\eta\{\xi_- - \eta\tilde{\alpha}(\xi_-\alpha_+ - \xi_+\alpha_-) + \eta\frac{\partial}{2cd\tau'}\tilde{\alpha}^2)\}, \quad (9)$$

and $z_o = (\xi_-\xi_+)^{-1}(\xi_+ + \xi_-)$ is a spatial shift of $A_-$ with respect to the envelope $A_+$ ($z_o \ll l_o$)

The coupled field $A_\sigma$ will move or stay together with the following amplitude ratio: $(g_-/\Omega_-)A_-(t,z) \cong (g_+/\Omega_+)A_+(t,z)$. This condition corresponds to the appearances of the so-called dark state in the double-Λ system [28], which can also dramatically change the interaction of the copropagating light fields. In comparison with the standing case, the group velocity $v = cM_1(\Omega_\Sigma^2/Ng_+^2)$ in Eq. (9) can be altered by changing the control fields, which is the third term of the $M_1$ in Eq. (9). However, the lowest velocity is only realized if $\left|\frac{\partial}{cd\tau'}\eta\tilde{\alpha}\right| \to 0$ and $(\xi_-\alpha_+ - \xi_+\alpha_-) \to 0$ occur with complete light stoppage ($v=0$). The parameters of the fields $A_+$ and $A_-$ are strongly coupled with each other: That means the field $A_-$ will be generated by the frequency shift $\omega_- = \omega_{41} + \delta\omega_+$ if the frequency of the field $A_+$ is tuned to $\omega_+ = \omega_{31} + \delta\omega_+$. The stationary field, $\hat{E}(t,z) \sim g_+^{-1}\Omega_+ \Psi_+ e^{-i\omega_+(t-z/c)} + g_-^{-1}\Omega_- \Psi_- e^{-i\omega_-(t+z/c)}$ is bound to the coherence moving grating that is completely different from the standing-wave grating in Ref. [20].

Taking into account the following condition $\gamma\gamma_2 << \Omega_\Sigma^2(t)$ in Eq. (9) we obtain a spatial size $l(t_s) = cB^2(t_s)$ of the two-color stationary light such that $l^2(t_s) = l_o^2 + \frac{1}{\xi_\Sigma}\int_0^{t_s} dt' v_b(t') + \delta X(t_s)$, where $v_b(t) = \left(\frac{2c}{Ng_\Sigma^2}\right)\frac{\Omega_+^2(t)\Omega_-^2(t)}{\Omega_\Sigma^2(t)}$ is an effective velocity (where $g_\Sigma^{-2} = g_+^{-2} + g_-^{-2}$ and $\xi_\Sigma^{-1} = \xi_+^{-1} + \xi_-^{-1}$) of the pulse spreading, $\delta X(t) = \xi_-^{-2}(\tilde{\alpha}^2(t) - \tilde{\alpha}^2(0))$ which is a small value. The broadening of the pulses exists only in the presence of the two control fields $\Omega_+$ and $\Omega_-$ and in the case of the zero group velocity, $\Omega_+/g_+ = \Omega_-/g_-$ we have $\delta l_b^2(t_s) = \xi_\Sigma^{-1} \int_0^{t_s} dt' v_b(t) = v_g t_s / \xi_\Sigma$. For this there are three possible cases: (1) $\delta l_b^2(t_s)\big|_{g_- << g_+} \cong \xi_-^{-1} v_{g+} t_s$ ; (2) $\delta l_b^2(t)\big|_{g_- = g_+} = 2\xi_+^{-1} v_g t_s$ and (3) $\delta l_b^2(t)\big|_{g_- >> g_+} = \xi_+^{-1} v_g t_s$. So the spatial spreading of the two-color light can be suppressed by choosing a particular medium satisfying $g_- \geq g_+$ and with a strong absorption coefficient $\xi_\sigma$.

For a single photon wave packet in the initial probe field, using solution (7) we found that the probability $P(\omega_+ \to \omega_-)$ of its transformation to the photon with a new carrier frequency $\omega_- = \omega_{41}$ is limited only by spatial broadening according to the following relation $P(\omega_+ \to \omega_-) = l_o / l(t_s) \cong \left(1 + l_o^{-2} \frac{v_b}{\xi_\Sigma} t_s\right)^{-1/2}$. If the stationary time satisfies the condition $t_s << v_g^{-1} l_o^2 \xi_+$ ($g_- \geq g_+$) we find that the initial field transforms to the new field with probability close to one: $P(\omega_+ \to \omega_-) \cong 1 - l_o^{-2} \frac{v_{g+}}{\xi_+} t \cong 1$. This process may also offer a new method of controlling the two spectral components of quantum fields in a coherent medium.

Note that the maximum storage time of the stationary lights is determined by the spin coherence decay time $\gamma_2^{-1}$ between the two lowest energy levels |1> and |2>. In the case of turning off both control laser pulses, the stationary lights disappear: it is the quantum mapping process discussed already in various experiments, in which the stationary quantum coherence is connected to the dark resonance of the levels $|1\rangle$ and $|2\rangle$ in the form of $P_{12} = N^{-1/2}(\alpha_+ \Psi_+ + \alpha_- \Psi_-)$ [9,16,21].

As stated in Eq. (7), we can initiate the two-color light to move together through the medium by breaking the stationary light condition by increasing one control field intensity. Such a movement of the two-color light resembles the properties of quantum entanglement in a traveling slow-light scheme in Ref. [10].

Let us now demonstrate the properties of the two-color stationary light different from the slow-light scheme [10] with an additional two-level system composed of M atoms localized in a spatial location close to r=R_a, (z=z_a) in a volume $V_a = S\delta l_a$ with a cross section S of the slow light fields and longitudinal size $\delta l_a << L$ (L is the spatial size of the medium). We assume the atomic frequency $\omega_a$ is close to that of the field $E_+$: $\omega_a - \omega_+ << \omega_a - \omega_-$. Ignoring the interaction of the field $E_-$ and taking into account the strong coupling of the fields $\Psi_-(t,z')$ and $\Psi_+(t,z)$ (see Eqs. (6) we find the equation for the field $\Psi_+(t,z)$ ignoring the small spatial spreading of the slow light:

$$\frac{\partial}{c\partial\tau}\Psi_+(\tau,r) + \frac{\xi_-\alpha_+ - \xi_+\alpha_-}{\xi_-}\frac{\partial}{\partial z}\Psi_+(\tau,r) = -\frac{(g_a^2/c)}{(\gamma_a + i(\omega_a - \omega_+))}\sum_{j=1}^M \delta(r - r_j)\Psi_+(\tau,r)$$ (10)

where $\gamma_a$ is effective decay rate for the transition, and $g_a$ is the coupling constant for photon–atom interaction. Assuming a constant group velocity v and $\gamma_a < \omega_a - \omega_+$, we obtain the following solution for the field after the interaction with this M atomic system: $\Psi_+(t, z > z_a) = \Psi_+(t, z < z_a)\exp\{i\varphi_t\}$, where $\varphi_t = \frac{M(g_a^2/c)}{(\frac{\xi_-\alpha_+ - \xi_+\alpha_-}{\xi_-})S(\omega_a - \omega_+)} = M\frac{\sigma_a \gamma_a}{(\frac{\xi_-\alpha_+ - \xi_+\alpha_-}{\xi_-})S(\omega_a - \omega_+)}$. Here the $\sigma_a$ is a cross section of the atom a. By contrast in the case of a usual slow-light scheme; $\alpha_+ = 1$ and $\alpha_- = 0$, an important condition of $\{\frac{\xi_-\alpha_+ - \xi_+\alpha_-}{\xi_-}\} = 1$ is obtained, which is independent of the initial group velocity of the light $v_g$. Thus the field gets additional phase $\varphi_e$ after nonresonant interactions with atoms, and this kind of phase increases has already been discussed in the slow-light scheme [6,10,19]. Here we note that the controllable and additional factor $\frac{\xi_-\alpha_+ - \xi_+\alpha_-}{\xi_-}$ determines the magnitude of the increased phase due to the enhancement of the interaction time. For a complete stationary light at t=$t_o$, we find the following solution of Eq. (10) $\Psi_+(t_s + t_o, z_a) \cong \Psi_+(t_o, z_a)\exp\{i\chi_s t_s\}$, where $\chi_s = M\frac{\sigma_a \gamma_a v_g}{S(\omega_a - \omega_+)\delta l_a}$. The solution is very close to the interactions of the photons with atoms in a cavity, where the maximum stopping time $t_s$ ($t_s = 1/\gamma_2$) of the two-color stationary light acts like a cavity decay time. Here it should be noted that the $t_s$ is much longer than interaction time obtained in the slow-light propagation scheme. This means that the present stationary-light scheme has advantages over the slow-light traveling scheme as it has no limitation on the spatial size of the nonlinear medium and the possibility of obtaining a π phase shift even with a single atom if $t_s > \delta l_a/v_g$ and S~$\sigma_a$. Thus, the stationary light in Eq. (12) results in enormous increase in the phase for π, and can principally be used for generation of macroscopic quantum entanglement, Schroedinger cat states, and quantum nondemonlition measurement as well as for the cavity-QED quantum optics.

In conclusion, we have presented the quantum manipulation of two-color stationary light using a backward nondegenerate four-wave mixing scheme in a double-Λ four-level system. In this scheme, a traveling quantum field can be quantum optically manipulated by simply adjusting the control fields' parameters for stationary light. We have also showed that the proposed two-color stationary light scheme includes the traveling light scheme with the advantage of enhanced nonlinearity. The proposed two-color stationary light scheme can also be a substitute for cavity-QED in quantum information processing by choosing an optically dense medium with long spin-decay time proportional to $\gamma_2^{-1}$. The two-color stationary light scheme, therefore, has potential in deterministic quantum information science with enhanced nonlinearity due to increased interaction time with a localized quantum system.

B.S. Ham acknowledges that this work was supported by Korea Research Foundation Grant KRF-2003-070-C00024.

FIG. 1. An energy level diagram of a double-Λ type four-level system for a two-color stationary light.

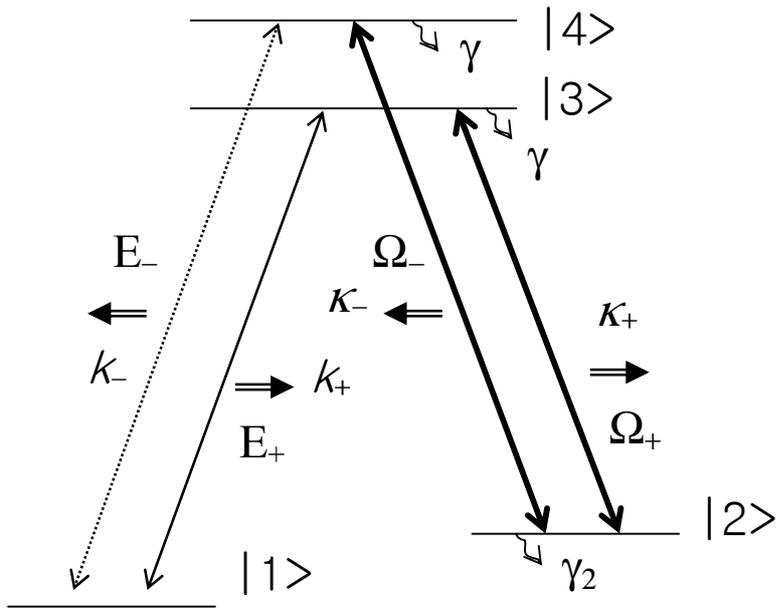